\documentclass[aps,pra,reprint,amsmath,amsfonts,amssymb,superscriptaddress]{revtex4-1}
\usepackage{graphicx,float,calc}
\usepackage{color,bm}
\usepackage{ulem}
\usepackage{braket}
\usepackage[colorlinks,urlcolor=blue,citecolor=blue,linkcolor=blue]{hyperref}
\begin{document}
\title{Topological Anderson insulators induced by random binary disorders}
\author{Shu-Na Liu}
\affiliation{Guangdong Provincial Key Laboratory of Quantum Engineering and Quantum Materials, School of Physics and Telecommunication Engineering, South China Normal University, Guangzhou 510006, China}
\author{Guo-Qing Zhang}\thanks{zhangptnoone@m.scnu.edu.cn}
\affiliation{Guangdong Provincial Key Laboratory of Quantum Engineering and Quantum Materials, School of Physics and Telecommunication Engineering, South China Normal University, Guangzhou 510006, China}
\affiliation{Guangdong-Hong Kong Joint Laboratory of Quantum Matter, Frontier Research Institute for Physics, South China Normal University, Guangzhou 510006, China}
\author{Ling-Zhi Tang}
\affiliation{Guangdong Provincial Key Laboratory of Quantum Engineering and Quantum Materials, School of Physics and Telecommunication Engineering, South China Normal University, Guangzhou 510006, China}
\author{Dan-Wei Zhang}
\affiliation{Guangdong Provincial Key Laboratory of Quantum Engineering and Quantum Materials, School of Physics and Telecommunication Engineering, South China Normal University, Guangzhou 510006, China}
\affiliation{Guangdong-Hong Kong Joint Laboratory of Quantum Matter, Frontier Research Institute for Physics, South China Normal University, Guangzhou 510006, China}

\begin{abstract}
Different disorders lead to various localization and topological phenomena in condensed matter and artificial systems. Here we study the topological and localization properties in one-dimensional Su-Schrieffer-Heeger model with spatially correlated random binary disorders. It is found that random binary disorders can induce the topological Anderson insulating phase from the trivial insulator in various parameter regions. The topological Anderson insulators are characterized by the disorder-averaged winding number and localized bulk states revealed by the inverse participation ratio in both real and momentum spaces. We show that the topological phase boundaries are consistent with the analytical results of the self-consistent Born approach and the localization length of zero-energy modes, and discuss how the bimodal probability affects the disorder-induced topological phases. The topological characters can be detected from the mean chiral displacement in atomic or photonic systems. Our work provides an extension of the topological Anderson insulators to the case of correlated disorders.
\end{abstract}
\date{\today}

\maketitle
\section{Introduction}
In the past decades, topological states of matter have been widely investigated in condensed-matter physics~\cite{Hasan2010,Qi2011,Bansil2016,Armitage2018,Chiu2016,Gao2018,Zhang2016,Qin2016} and artificial systems such as ultracold atoms \cite {Zhang2018,Cooper2019,Goldman2016}, superconducting circuits \cite{Schroer2014,Roushan2014,XTan2018,XTan2019a,XTan2019b}, and photonic lattices \cite{LLu2014,Ozawa2019}. Under strong disorders, the topological states usually become trivial Anderson insulators due to the fact that the eigenstates become localized and the band gap  closes~\cite{Anderson1958,Klitzing1980,Evers2008}. Surprisingly, it was found that topological insulators can be induced by random disorders from trivial phases~\cite{Li2009}, leading to the novel interplay effect of topology and disorder. The disorder-induced topological phases are dubbed topological Anderson insulators (TAIs) and have been widely investigated~\cite{Li2009,Groth2009,Jiang2009,SanchezPalencia2010,Meier2018,Li2020,MondragonShem2014,Hsu2020,Jiang2019,Zhang2020,Tang2020,Titum2015,Yang2021,Zhang2021,Stutzer2018,Peng2021,Zhang2013,Shi2021,Liu2020,Hsu2020,Liu2020a,Liu2021,Zhang2012,Song2012,Xu2012,Zhang2019,Yamakage2013,Orth2016,Xing2011,Shen2017,Prodan2011,Wu2019,Altland2014,Longhi2020}. Experimental observations of TAIs were reported in several artificial systems, such as 1D cold atomic wires~\cite{Meier2018}, 2D photonic waveguide arrays~\cite{Stutzer2018,Liu2020}, electric circuits~\cite{Zhang2021,Zhang2019}, coupled waveguides~\cite{Stuetzer2015}, and photonic quantum walks~\cite{Lin2021}. {\color{black}Notably, TAIs are induced by the uncorrelated disorders in these studies. For instance, both uncorrelated random~\cite{Meier2018} and quasiperiodic disorders~\cite{Zhang2021a,Tang2022,Lu2022} can give rise to the TAI phase in 1D Su-Schrieffer-Heeger (SSH) chains. It was shown that the spatial correlations in the disorder can entirely suppress the emergence of the TAI phase on two-dimensional quantum spin Hall systems with Gaussian correlated on-site potentials~\cite{Girschik2013}. However, the effects of correlated disorders on other topological phases are largely unexplored. In particular, it remains unclear whether the TAI phase exhibits in 1D SSH chiral chains with correlated disorders.}

As a typical kind of spatially short-range correlated disorder, the random binary disorder has been studied in localization phenomena. It was found that the random binary disorder can induce unconventional localization-delocalization transition~\cite{Flores1989,Hilke1997,Dunlap1990,Sedrakyan2004,Schaff2010}, and the phase transitions between Mott insulators and Bose glasses ~\cite{Krutitsky2008} or metal phases~\cite{Semmler2010}. Moreover, the random site energies with a discrete binary probability distribution can significantly influence the electronic structure and transport properties~\cite{Evangelou1993,Alvermann2005}. The random binary disorder in random dimer models can be used to explain the insulator-metal transition with new metallic states in conducting polymers~\cite{PHILLIPS1991} and a new transport mechanism~\cite{Wu_1991}. Realizations of the random-dimer model have been achieved in GaAs-AlGaAs superlattices~\cite{Bellani1999}, 1D waveguide array fabricated infused silica~\cite{Naether_2013}, and a ultracold atomic mixture in 1D optical lattices~\cite{Schaff2010}. \textcolor{black}{Except for these localization effects, it is interesting to study the topological phases under the random binary disorder. A nature question is whether the TAI phase can be induced by this kind of correlated disorder.}

In this paper, we study the topological and localization properties of the 1D SSH model ~\cite{Su1979,Heeger1988} with random binary disordered hoppings. We obtain the topological phase diagrams by numerically calculating the real-space winding number, the energy gap, and the edge states. It is found that the random binary disorder can induce the TAI phase from the trivial insulator in a wide range of parameter regions. We also study the influence of the bimodal probability on the topological phase. By generalizing the existing self-consistent Born approximation (SCBA) approach to our model, we reveal the trivial-topological phase transition driven by the disorder-renormalization of the intracell hopping strength. An analytical formula for the localization length of the zero-energy state is derived, whose divergence indicates the topological phase transition. We further study the localization properties of the system by computing the inverse participation ratio (IPR) in both real and momentum space. We numerically show that the mean chiral displacement of the bulk dynamics can be used to reveal the topological phases, which can already be implemented in ultracold atomic and photonic systems~\cite{Cardano2017,Maffei2018,Meier2018,Xiao2021,Wang2018,DErrico2020,Xie2020,Xie2019,Wang2021,Meng2020}.
Finally, we discuss the TAI phase in a more general case with both binary disordered inter- and intra-cell hoppings.

The rest of this paper is organized as follows. We first introduce the 1D SSH model with random binary disordered hoppings in Sec.~\ref{sec2}. Section~\ref{sec3} is denoted to present the results of the topological phase diagrams with disorder-induced TAIs, the localization properties, and the influence of the bimodal probability. In Sec.~\ref{sec4}, we propose to detect the topological phases by measuring the mean chiral displacement. A short discussion and a brief summary are finally given in Sec.~\ref{sec5}.

\section{\label{sec2}Model}
We begin by considering the 1D SSH model with random binary disordered hopping strengths. The tight-binding Hamiltonian consists of two sublattices (labeled by A and B, respectively) per unit cell, which can be expressed as
\begin{equation}\label{H}
{H}=\sum_{n} (m_{n} c_{A, n}^{\dagger} c_{B,n}
+ \mathrm{h.c.})+\sum_{n}( t_{n} c_{A, n+1}^{\dagger} c_{B, n}+\mathrm{h.c.}),
\end{equation}
where $c_{A(B), n}^{\dagger}$ and $c_{A(B), n}$ are the creation and annihilation operators of electron on $A(B)$ sublattice in the $n$-th unit cell, respectively. We consider a system has $N$ unit cells, and the total number of sublattice sites is $L=2N$ with each sublattice site indexed by $x\in [1,L]$.

\textcolor{black}{For simplicity, we first consider the dilute case where the intercell hopping strengths to be site-independent $t_n=t$, and the intracell hopping strengths $m_n$ are binary disorders. In the last section, we will generalize our results by considering $t_n$ also binary disordered.} To do this, $m_n$ is drawn from a bimodal probability distribution function~\cite{Saha1996,Mondal2019}
\begin{equation}
\mathcal{P}(m_{n})=P \delta(m_{n}-V_0-w_a) + (1-P) \delta(m_{n}-V_0-w_b),
\end{equation}
which means $m_n$ takes bimodal value $V_0+w_a$ or $V_0+w_b$ with probability $P$ or $1-P$ on $n$-th unit cell. $V_{0}=w_{a}=1$ is chosen as the energy unit, and the disorder strength is defined as the difference between $w_a$ and $w_b$: $W\equiv w_a-w_b$. For the special case that the probability $P=0$ or $P=1$, it reduces to the clean SSH model~\cite{Su1979,Heeger1988} of free fermions.

It is well-known that the SSH model has two topologically distinct phases and exists two-fold degenerate zero-energy edge states ~\cite{Asboth2016} in the topological insulating phase under open boundary conditions (OBCs). In experiments, the SSH model has been realized in different systems, such as cold atoms in 1D optical superlattices~\cite{Atala2013}. In this clean case, the 1D winding number of Bloch vectors in momentum space can be used to characterize topologically trivial and nontrivial phases. In the presence of disorders, the translational symmetry breaking and the wave vectors in momentum space are no longer good quantum numbers, so the conventional winding number defined in terms of Bloch vectors should be extended to the real space~\cite{MondragonShem2014}. Here we use the real-space winding number to characterize the topological property of our model. For a given disorder configuration denoted by the label $s$, the corresponding real-space winding number is given by ~\cite{MondragonShem2014}:
	\begin{equation}
		\nu_s=\frac{1}{L^{\prime}} \operatorname{Tr}^{\prime}(\Gamma Q_s[Q_s, X]),
	\end{equation}
	where $X=\mathrm{diag}(1,1,2,2,\cdots,N,N)$ denotes the unit-cell coordinate operator, $Q_s=\sum_{j}(\ket{\psi_{j}}_{ss}\bra{\psi_{j}}-\ket{\tilde{\psi}_{j}}_{ss}\bra{\tilde{\psi}_{j}})$ is the flat-band Hamiltonian with $\ket{\psi_j}_s$ being the $j$-th eigenstate of $H$ and $\ket{\tilde{\psi}_{j}}_{s}=\Gamma^{-1}\ket{\psi_j}_s$, $\Gamma=\mathrm{diag}(1,-1,1,-1,\cdots,1,-1)$ is the chiral operator with diagonal elements $1$ and $-1$ for A and B sublattices, respectively. Here $L^{\prime}=L-2l$ is slightly less than the chain length $L$ to get rid of the $l$ sites boundary effect at both ends, and ${\mathrm{Tr}}^{\prime}$ stands for the trace within the central region of the length $L^{\prime}$ in the chain. \textcolor{black}{In numerical simulations, we choose $l=L/4$ to avoid boundary effects and get quantized winding numbers.} For the disordered SSH model, we use the disorder-averaged (over $N_s$ configurations) winding number
\begin{equation}\label{winding number}
\nu=\frac{1}{N_{s}} \sum_{s=1}^{N_{s}} \nu_{s}
\end{equation}
as the topological invariant, where a modest configuration number $N_{s}$ is usually sufficient in practice.
Thus, $\nu=0$ and $1$ correspond to topologically trivial and nontrivial in the limit of $L \rightarrow \infty$, respectively.

\section{\label{sec3}Topology and localization}
In this section, we systematically investigate the topological and localization properties of the 1D SSH model with random binary disordered hoppings. First, we study the winding number, energy spectrum, and edged state to consistently characterize the topological phase transition induced by random binary hoppings. This phase transition can also be revealed by using the SCBA method which accounts for the disorder effects into the renormalization of the hopping strengths. We then analytically derive the localization length of the zero-energy edge mode and numerically calculate the real-space and momentum-space IPR. Finally, we study the influence of bimodal probability on the topological and localization properties.

\begin{figure}[tb]
	\includegraphics[width=0.48\textwidth]{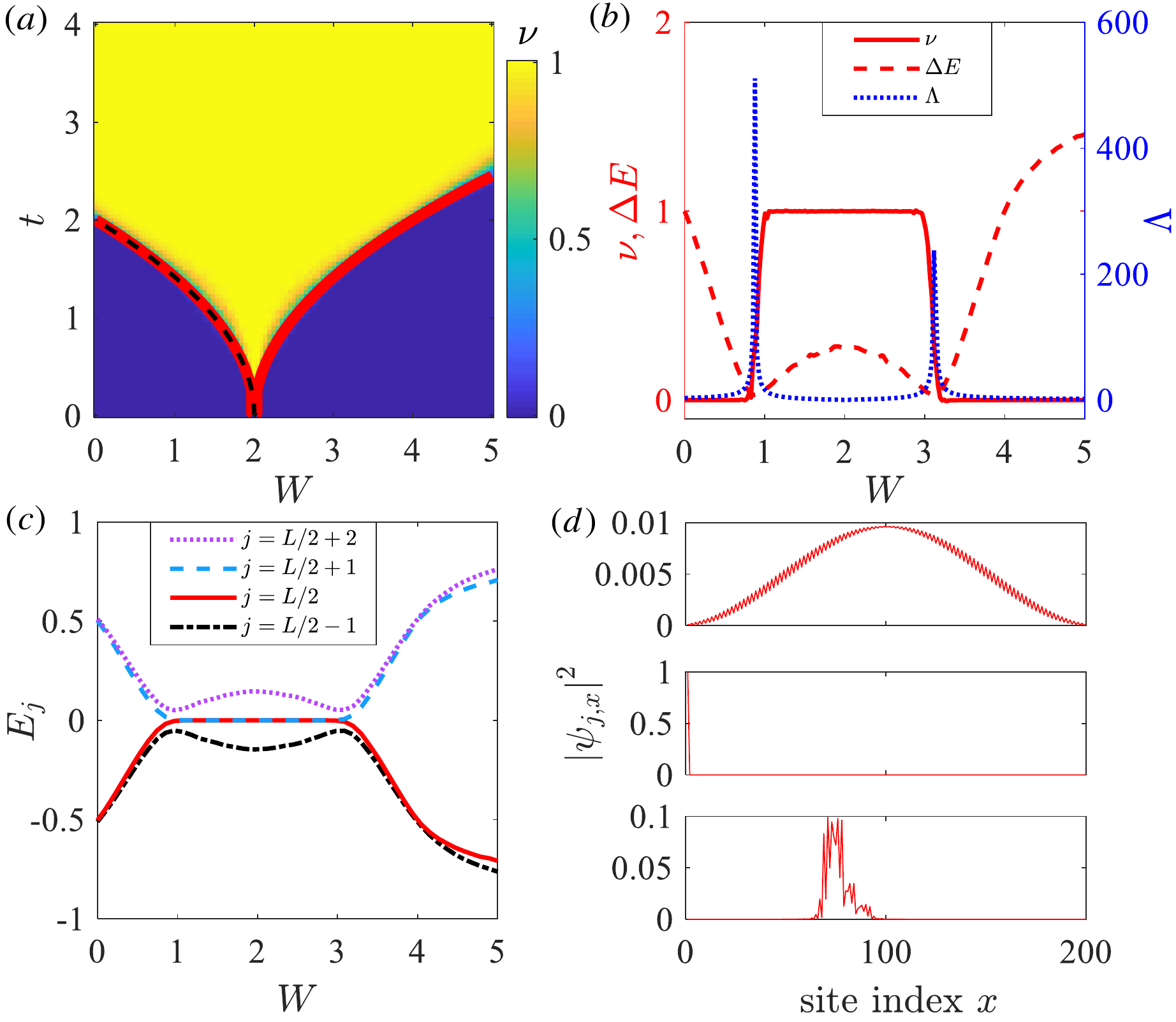}
	\caption{(Color online) (a) Disorder-averaged real-space winding number $\nu$ in the $W$-$t$ plane. The black dashed line denotes the topological phase boundary determined by the SCBA method. The solid red line is the topological phase boundary revealed by divergence of the localization length $\Lambda$ of zero energy modes. (b) The winding number $\nu$, the energy gap $\Delta E$, and the localization length $\Lambda$ as a function of $W$ for $t=1.5$. (c) Four centered eigenenergies under OBCs as a function of $W$ for $t=1.5$. (d) Density distribution of the $j=L/2$-th eigenstate for $t=1.5$ and $W =0,2,5.3$ (from top to bottom), respectively. Other parameters are $P=1/2$, and $L=200$.}
	\label{fig1}
\end{figure}

\subsection{\label{sec3.1}Topological phase diagrams}
We first investigate the topological properties of the 1D random binary disordered SSH model with fixed $P=1/2$. By calculating the disorder-averaged real-space winding number $\nu$ for the system of $L=200$ on the $W$-$t$ plane, we obtain the topological phase diagram shown in Fig.~\ref{fig1}(a). In the clean limit $W=0$, the system is in the trivial phase with $\nu=0$ for intercell hopping $t<2$. Interestingly, there exists nontrivial $\nu=1$ phase in the $t<2$ region with moderate disorder strength, indicating the existence of TAIs induced by the random binary disorder. We also compute the bulk gap under PBCs and the density distribution of the most centered two eigenstates under OBCs. In this case, the disorder-averaged $j$-th eigenenergy $E_j$ and the corresponding bulk gap $\Delta E$ are given by
\begin{align}
			E_j&=\frac{1}{N_s}\sum_{s=1}^{N_s}E_{s,j},\\
			\Delta E&=\frac{1}{N_{s}} \sum_{s=1}^{N_{s}}\left|E_{s,L / 2+1}-E_{s,L / 2}\right|.
\end{align}
In Fig.~\ref{fig1} (b), we plot $\nu$ and $\Delta E$ as functions of $W$ for $t=1.5$. Our results show that the topological phase transition points revealed by the abrupt changes of real-space winding number $\nu$ under OBCs and closure points of bulk gap $\Delta E$ under PBCs agree with each other. In addition, there are two topological phase transitions including one from trivial phase to nontrivial phases for $W \simeq 0.88$ and another from nontrivial to trivial phases for $W \simeq 3.12$, respectively. In the TAI phase when $0.88\lesssim W\lesssim3.12$ under OBCs, two-fold degenerate zero-energy modes exhibit in the energy spectrum shown in Fig.~\ref{fig1}(c). The corresponding density distributions of the $j=L/2$-th eigenstate for three typical disorder strengths $W = 0,2,5.3$ are plotted in Fig.~\ref{fig1}(d). The system in the clean limit is in the trivial band insulator with the extended bulk states at the center of the energy spectrum. In the TAI phase under the moderate disorder strength, the two centered eigenstates become zero-energy edge mode localized at two ends. For sufficient disorder strength, the system becomes a trivial Anderson insulator where the eigenstates are localized in the bulk.

Now we use the SCBA analysis to explain the disorder-induced trivial-topological phase transition. The disorder effect can be renormalized to the parameters of the clean Hamiltonian based on an effective medium theory and the SCBA method~\cite{{Groth2009,Chen2015,Park2017,Liu2016}}. The key ingredient is to self-consistently obtain the self-energy caused by the disorder, which acts as the renormalization of the clean Hamiltonian. For our model, the self-energy term $\sum(E_F)$ satisfies the following self-consistent equation
	\begin{equation}
		\frac{1}{E_{F}-H_{q}(k)-\Sigma\left(E_{F}\right)}=\left\langle\frac{1}{E_{F}-H_{\text{eff}}(k,W)}\right\rangle_s.
	\end{equation}
Here $\braket{\cdots}_s$ means taking average over sufficient disorder configurations, $H_{q}(k)$ is the clean Hamiltonian ($W=0$) in momentum space with intracell hopping strengths $m=V_0+w_a$, $E_{F}\equiv0$ is the Fermi energy, and the effective binary-disordered Hamiltonian is given by
\begin{align}\label{Hkn}
		H_{\mathrm{eff}}(k,W)=(m'+t \cos k) \sigma_{x}+t\sin k \sigma_{y},
\end{align}
where $m'=m$ with probability $P$, and $m'=m-W$ with probability $1-P$. The self-consistent self-energy equation under the first-order SCBA can be written as~\cite{gonccalves2018haldane}
	\begin{equation}\label{scbabz}
		\Sigma=\frac{\sigma^2}{2\pi}\int_0^{2\pi}\left\langle\frac{1}{G_0^{-1}-\Sigma}\right\rangle_s dk,
	\end{equation}
where $G_0^{-1}=-H_{\mathrm{eff}}(k,W)$ is the Green's function, and $\sigma^2=W^2/4$ is the variance of the binary disorder distribution. The self energy can be parameterized as $\Sigma=\Sigma_\mathrm{I}I+\Sigma_x \sigma_x+\Sigma_y \sigma_y+\Sigma_z \sigma_z.$
	As the inverse of Hamiltonians (\ref{Hkn}) has only $x$ and $y$ components, so the self energy $\Sigma_z=0$ can satisfy Eq.~(\ref{scbabz}). The $\Sigma_y$ component from the off-diagonal entries of $G_0$ can be explicitly calculated as $\Sigma_y=\int_0^{2\pi}\mathrm{Im}[\braket{G_0}_s] dk=0$. Finally, we obtain the non-vanishing $\Sigma_x$ component, which renormalizes the intracell hopping strength as
	\begin{equation}\label{eq_crit}
		\bar{m}=m+\Sigma_{x}(W).
	\end{equation}
In the clean limit, $\bar{m}=2$ and the phase transition point occurs at $\bar{m}=t=2$. For the weak disorder $W \leq 2$ and under the SCBA, the renormalization of the hopping strength indicates the topological phase transition happens at the critical value $W_c$ where $\bar{m}=t$. We numerically obtain the self energy $\Sigma_{x}(W)$ and determine the topological phase transition boundary for weak disorder strength. The SCBA method fails at large disorders for $W>2$. We plot the left-hand-side phase boundary revealed by this method as the black dashed line in Fig.~\ref{fig1}(a), which is consistent with the topological transition boundary revealed by the winding number.

\subsection{\label{sec3.2}Localization properties}

Now we show that the divergence of the localization length of zero-energy states is related to the topological phase transition due to the appearance of exponentially localized zero-energy edge modes. The Schr\"odinger equation for the zero-energy eigenstate $\psi$ in this case reads $\hat{H} \psi=0$, which leads to $m_{n} \psi_{n, A}+t_{n} \psi_{n+1, A}=0$ and $t_{n} \psi_{n, B}+m_{n+1} \psi_{n+1, B}=0$. We can obtain the following solutions
	\begin{align}
		\psi_{N, A}&=(-1)^{N} \prod_{n=1}^{N} \frac{m_{n}}{t_{n}} \psi_{1, A},
\\
		\psi_{N, B}&=(-1)^{N} \prod_{n=1}^{N} \frac{t_{n}}{m_{n+1}} \psi_{1, B}.
	\end{align}
	The inverse of the localization length is then given by (in the thermodynamic limit $N \rightarrow \infty$)~\cite{MondragonShem2014}
	\begin{equation}\label{appeq}
		\Lambda^{-1}=\max \{\lim _{N \rightarrow \infty} \frac{1}{N} \ln{|\psi_{N, A}|}, \lim _{N \rightarrow \infty} \frac{1}{N} \ln{|\psi_{N, B}|}\}.
	\end{equation}
	Taking $\psi_{1, A}=\psi_{1, B}=1$, we can obtain
		\begin{align}\label{appeql}
			\lim _{N \rightarrow \infty} \frac{1}{N} \ln{ |\psi_{N, A}|}
			&=\lim _{N \rightarrow \infty} \frac{1}{N} \ln{ |\psi_{N, B}|} \\
			&=|\lim _{N \rightarrow \infty} \frac{1}{N} \sum_{n=1}^{N}(\ln{|t_{n}|}-\ln{|m_{n}})|.\nonumber
		\end{align}
Inserting Eq.~(\ref{appeql}) into Eq.~(\ref{appeq}), we obtain inverse of the localization length for the zero-energy mode as
	\begin{equation}
		\Lambda^{-1}=|\lim _{N \rightarrow \infty} \frac{1}{N} \sum_{n=1}^{N}(\ln{|t_{n}|}-\ln{|m_{n}|})|.
	\end{equation}
	For our random binary disordered SSH model with $t_{n}=t$ and $m_{n}=V_{0}+w_{n}$, we can obtain the analytical result
	\begin{equation}\label{appeq2}
			\Lambda^{-1}=|\ln | t|-P\ln (2)-(1-P)\ln |2-W||,
	\end{equation}
when $V_0=1$, and $W=w_a-w_b$. We plot $\Lambda^{-1}=0$ as the red solid line in Fig.~\ref{fig1} (a), which shows the topological phase boundary can be revealed by the divergence of $\Lambda$. To be more clear, we also plot $\Lambda$ as the blue solid line in Fig.~\ref{fig1}(b). These results indicate that the topological phase transition from the trivial band insulator to the TAI in this disordered system can be revealed by the winding number, the energy gap, the renormalization under the SCBA method, and the divergence of localization length.

\begin{figure}[tb]
	\includegraphics[width=0.48\textwidth]{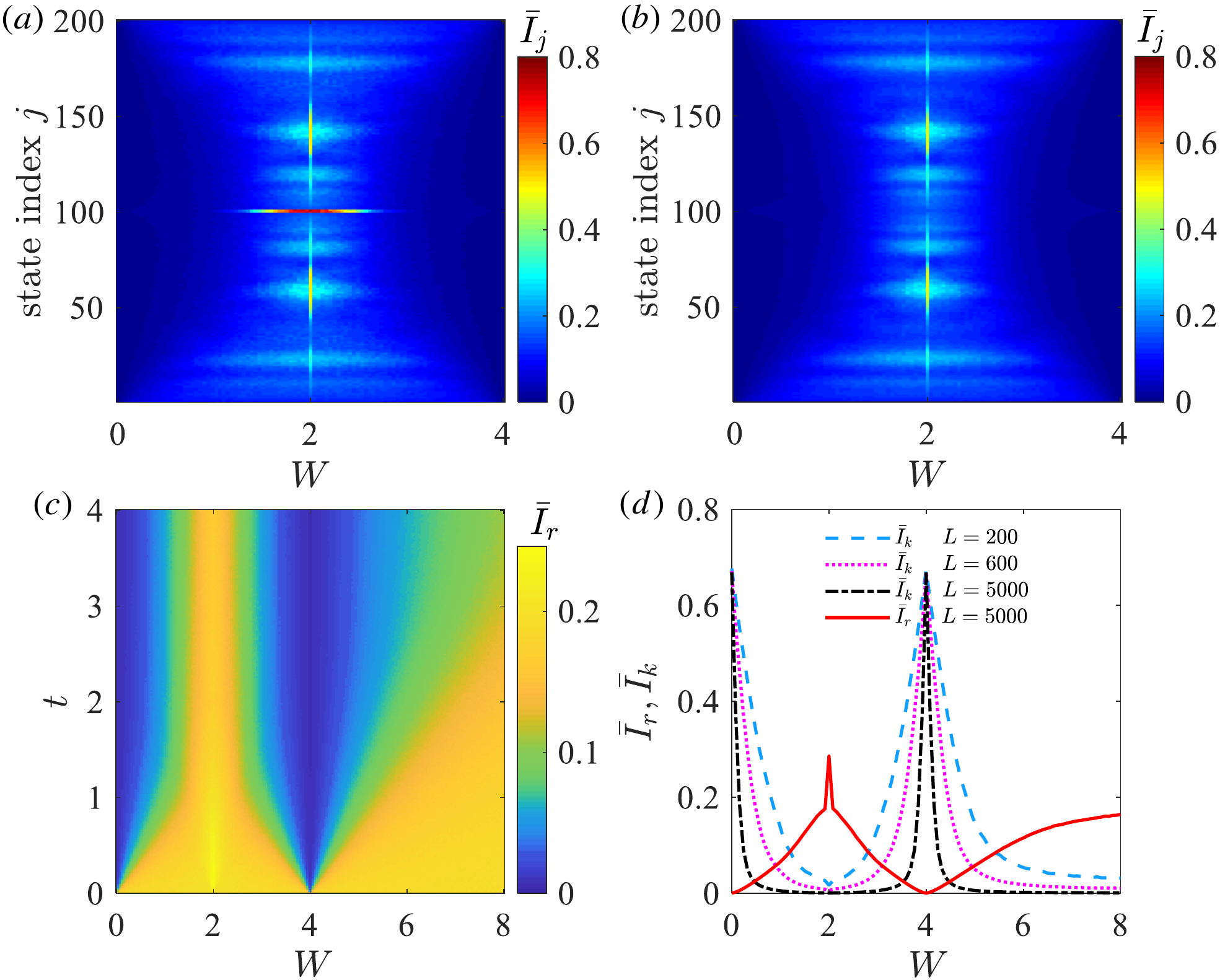}
	\caption{(Color online) \textcolor{black}{The disorder-averages IPR of each eigenstate for $L=200$ and $t=1.5$ under OBCs (a) and PBCs (b), respectively.} (c) The disorder- and eigenstate-averaged IPR $\bar{I}_{r}$ in the real space as functions of $W$ and $t$ for $L=200$ under PBCs. (d) The momentum (real) space IPR ${\bar{I}_{k}}$ ($\bar{I}_{r}$) as a function of $W$ for $t=1.5$ and various $L$s.}\label{fig2}
\end{figure}

We further investigate the bulk states localization induced by the random binary disorder by computing the IPR in both real and momentum space. \textcolor{black}{The disorder-averaged real-space IPR of the $j$-th eigenstate reads
\begin{equation}
	\bar{I}_{j}=\frac{1}{N_{s}} \sum_{s=1}^{N_{s}} \sum_{x=1}^{L}\left|\psi_{j, x}^{(s)}\right|^{4}
\end{equation}
where $\psi_{j,x}^{(s)}$ denotes the probability amplitude of the $j$-th normalized eigenstate at the $x$-th lattice site for the disorder configuration $s$. We present $\bar{I}_{j}$ for $L=200$ and $t=1.5$ under OBCs and PBCs in Figs.~\ref{fig2}(a) and (b). The IPRs of the centered two eigenstates ($j=100, 101$) show a large value in Fig.~\ref{fig2}(a), indicating the existence of two zero-energy edge states in the topological region under OBCs. The difference of IPRs between OBCs and PBCs is merely at the centered two states in the topological region. Therefore, we can use IPR under PBCs to characterize the localization of bulk states without topological edge modes.}

To characterize the bulk localization, the disorder- and eigenstate-average IPR in the real space under PBCs is given by
\begin{equation}
	\bar{I}_{r}=\frac{1}{N_{s}} \frac{1}{L} \sum_{s=1}^{N_{s}} \sum_{j=1}^{L} \sum_{x=1}^{L}\left|\psi_{j, x}^{(s)}\right|^{4}
\end{equation}
For extended state, real-space IPR $\bar{I}_{r}\sim 1/L$ tends to $0$ in the thermodynamic limit $L\rightarrow\infty$, while $\bar{I}_{r}\sim\mathcal{O}(1)$ for localized states. The numerical result of $\bar{I_r}$ in the $W$-$t$ plane for system size $L=200$ is shown in Fig.~\ref{fig2}(c), where both extended and localized phases can be clearly seen. We also consider the momentum-space IPR, and the disorder- and eigenstate-average IPR in the momentum space is given by
\begin{equation}
	\bar{I}_{k}=\frac{1}{N_{s}} \frac{1}{L} \sum_{s=1}^{N_{s}} \sum_{j=1}^{L} \sum_{k_{d}}^{2\pi}\left|\psi_{j, k_{d}}^{(s)}\right|^{4}
\end{equation}
The probability amplitude $\psi_{j,k_{d}}^{(s)}$ is obtained by taking the Fourier transformation $\psi_{j, k_{d}}^{(s)}=\frac{1}{L} \sum_{x=1}^{L} e^{-i k_{d} \cdot x} \psi_{j, x}^{(s)}$ where the wave vector is $k_{d}=2 \pi d / L$ with integer $d\in [1,L]$. A localized state in the real space becomes extended in the momentum space, and thus $\bar{I}_{k}\rightarrow 0$ in the large $L$ limit indicates localized phase. The results of $\bar{I}_{k}$ and $\bar{I}_{r}$ as functions of $W$ for $t=1.5$ are shown in Fig.~\ref{fig2}(d). Except those near $W=0,4$, $\bar{I}_{k}$ can approach to zero by increasing the lattice length $L$, which indicates the localized phase. For $W=0,4$, the vanishing of real-space $\bar{I}_{r}$ and large value of momentum-space $\bar{I}_{k}$ indicate the fully extended phase.

	\begin{figure}[tb]
		\includegraphics[width=0.48\textwidth]{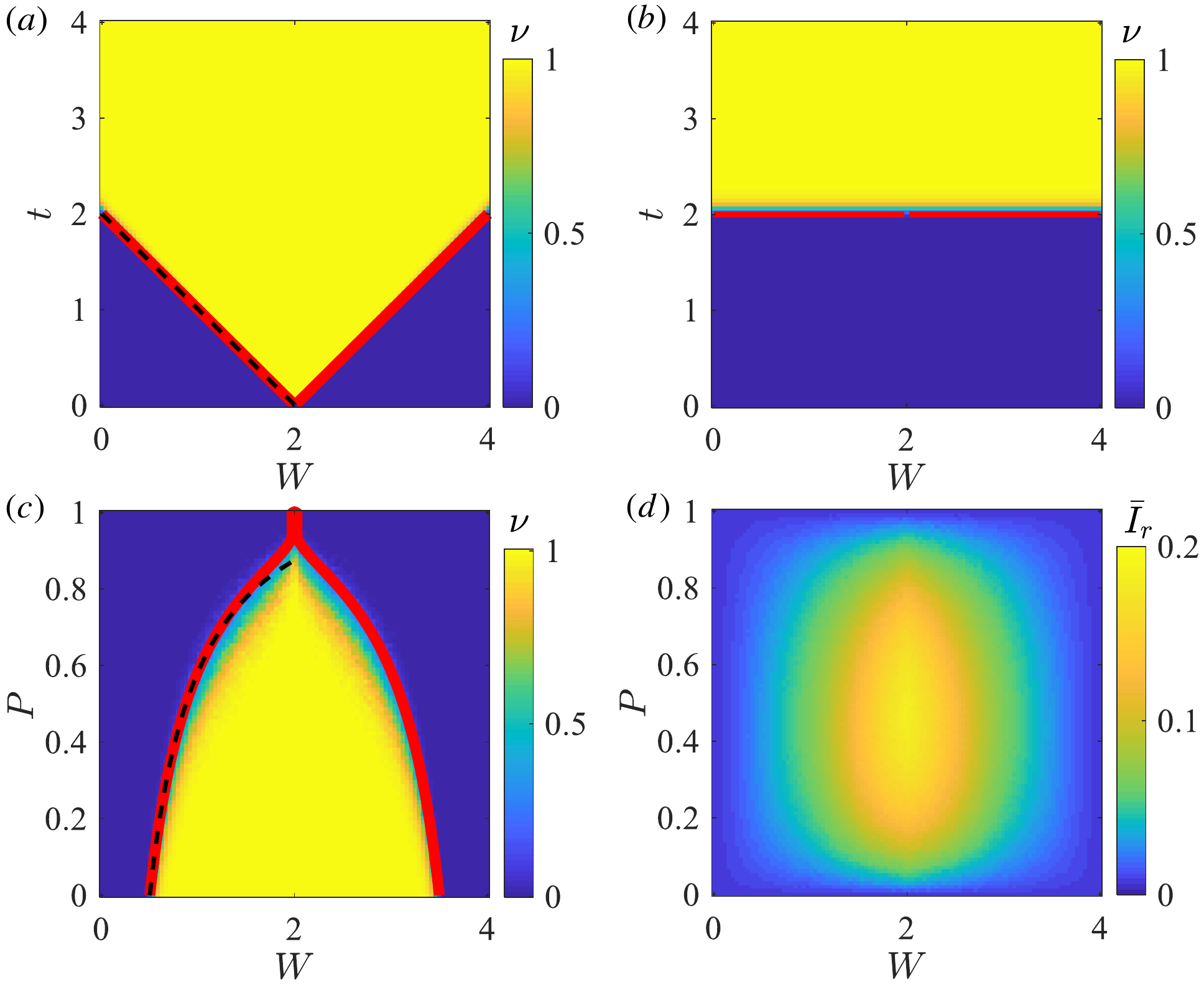}
		\caption{(Color online) Top panel: The topological phase diagram for $P=0$ (a) and $P=1$ (b) of $L=200$ systems revealed by the disorder-averaged winding number $\nu$ in the $W$-$t$ plane. Bottom panel: The winding number $\nu$ (c) and real-space IPR $\bar{I}_{r}$ (d) as functions of $P$ and $W$ for $t=1.5$ in $L=200$ systems. Solid red curves and dashed black lines indicate the topological phase boundaries by the divergence of localization length $\Lambda$s and by the SCBA method, respectively.
		}\label{fig3}
	\end{figure}

\subsection{\label{sec3.3} The influence of the bimodal probability}	

\textcolor{black}{It has been shown that the bimodal probability has the influence on the localization lengths in binary disordered systems \cite{Dunlap1990, Evangelou1993}. Now we consider how the phase diagram and localization properties are affected by the bimodal probability $P$ in the SSH model.} We first consider the topological properties in two limits of $P=0$ and $P=1$. The corresponding topological phase diagrams on the $W$-$t$ plane are plotted in Figs.~\ref{fig3}(a) and ~\ref{fig3}(b), respectively. For $P=0$ in Fig.~\ref{fig3}(a), we can see that there exist disorder-driven TAIs for $t$ varying from $0$ to $2$. The parameter region of the TAI phase reduces when decreasing $t$. For $P=1$ in Fig.~\ref{fig3}(b), the phase diagram is divided into topological trivial and nontrivial regions by the horizontal line $t=2$, which indicates the absence of disorder-induced TAIs. In Fig.~\ref{fig3} (c), we show the topological phase diagram on the $W$-$P$ plane for $t=1.5$. The parameter region of TAIs shrinks to zero as $P$ increases to unity. The solid red curves in Figs.~\ref{fig3}(a-c) correspond to the critical case of $\Lambda^{-1} \rightarrow 0$. The dashed black lines in Figs.~\ref{fig3}(a) and \ref{fig3}(c) are phase boundaries revealed by the renormalized hopping term $\bar{m}=t$ in the SCBA calculation. The corresponding real-space IPR $\bar{I}_{r}$ as functions of $W$ and $P$ is plotted in Fig.~\ref{fig3}(d), which shows that the system is more localized for moderate $W$ and $P$. \textcolor{black}{Therefore, by increasing $P$, there exist the topological-trivial phase transition and the localization-delocalization transition in the SSH model.}
	
\begin{figure}[tb]
	\includegraphics[width=0.45\textwidth]{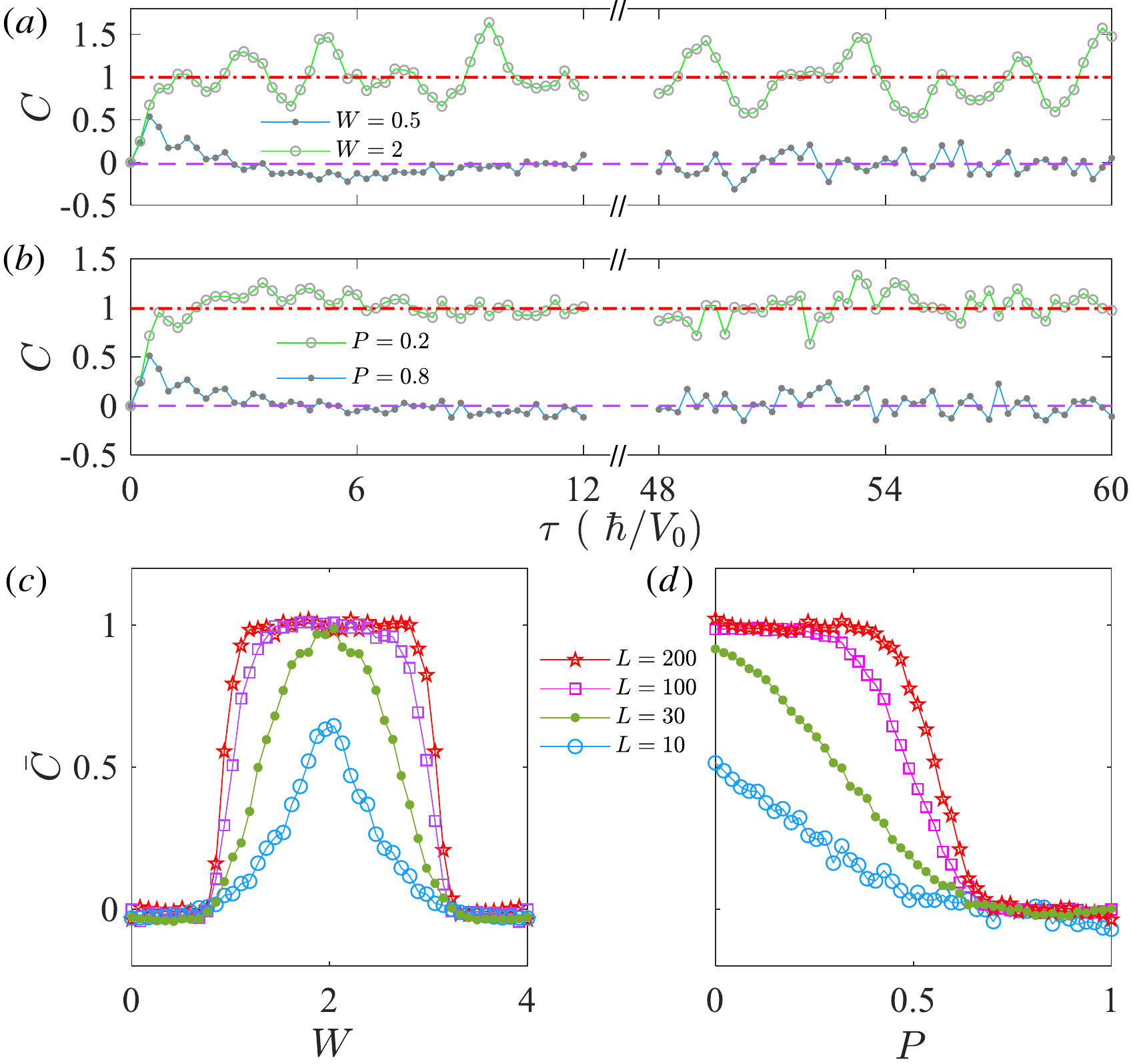}
	\caption{(Color online) The time evolution of $C$ for $P=1/2$ (a) and $W=1$ (b) with $t=1.5$ and $L=100$ under OBCs. The time- and disorder-averaged $\bar{C}$ as a function of $W$ for $P=1/2$ (c) and as a function of $P$ for $W=1$ (d) for $L=10,30,100,200$. The purple dashed line and red dot-dashed lines in (a) and (b) correspond to the values of $\bar{C}\approx\nu=0,1$ for topological and trivial phases, respectively.
	}\label{fig4}
\end{figure}

\section{\label{sec4}Proposed detections}
Finally, we propose that the TAI phase in this binary disordered model can be detected from the mean chiral displacement, as recently demonstrated in atomic and photonic systems~\cite{Cardano2017,Maffei2018,Meier2018,Xiao2021,Wang2018,DErrico2020,Xie2020,Xie2019,Wang2021,Meng2020, Xu2018,StJean2021}. The time evolution of the mean chiral displacement $C_{s}(\tau)$ for a single disorder configuration $s$ is given by
\begin{align}
	C_{s}(\tau)=2_{s}\langle\psi(\tau)|\Gamma X| \psi(\tau)\rangle_{s},
\end{align}
where $\tau$ denotes time, $\Gamma$ and $X$ are chiral and unit-cell coordinate operators previously used in the real-space winding number. The disorder average mean chiral displacement is then written as
\begin{align}
	C(\tau)=\frac{1}{N_{s}} \sum_{s=1}^{N_{s}}C_{s}(\tau).
\end{align}
After a long time evolution $\tau=T$, the mean chiral displacement averaged over disorders and time is given by
\begin{align}
	\bar{C}=\frac{1}{T}\int_{0}^{T} C(\tau) d\tau.
\end{align}
We choose the initial state at the center of the lattice and simulate the quantum dynamics for many disorder configurations, with typical results of $C(\tau)$ and $\bar{C}$ shown in Fig.~\ref{fig4}. As shown in Figs.~\ref{fig4}(a) and \ref{fig4}(b), the dynamics of $C(\tau)$ shows a transient and oscillatory behavior and its time- and disorder-averaged $\bar{C}$ converges to the winding number $\nu$ in both clean and disorder cases in the long time limit~\cite{Cardano2017,Maffei2018}.

We further plot $\bar{C}$ as a function of $W$ for $P=1/2$ and $t=1.5$ in Fig.~\ref{fig4}(c) for different lattice sizes. We can see that $\bar{C}$ converges to an integer value which approaches to $0$ in the trivial phase and $1$ in the topological phase by increasing the lattice size $L$, which agrees with the result of $\nu$ shown in Fig.~\ref{fig1}(b). We also numerically calculate $\bar{C}$ as a function of $P$ for $W=1$ and $t=1.5$ in Fig.~\ref{fig4}(d), which shows that $\bar{C}\approx1$ for small $P$ and drops to zero for $P \gtrsim 0.5$. The decrement of $\bar{C}$ by increasing $P$ is in good agreement with the winding number $\nu$ obtained in Fig.~\ref{fig3}(c).

\begin{figure}[tb]
	\includegraphics[width=0.45\textwidth]{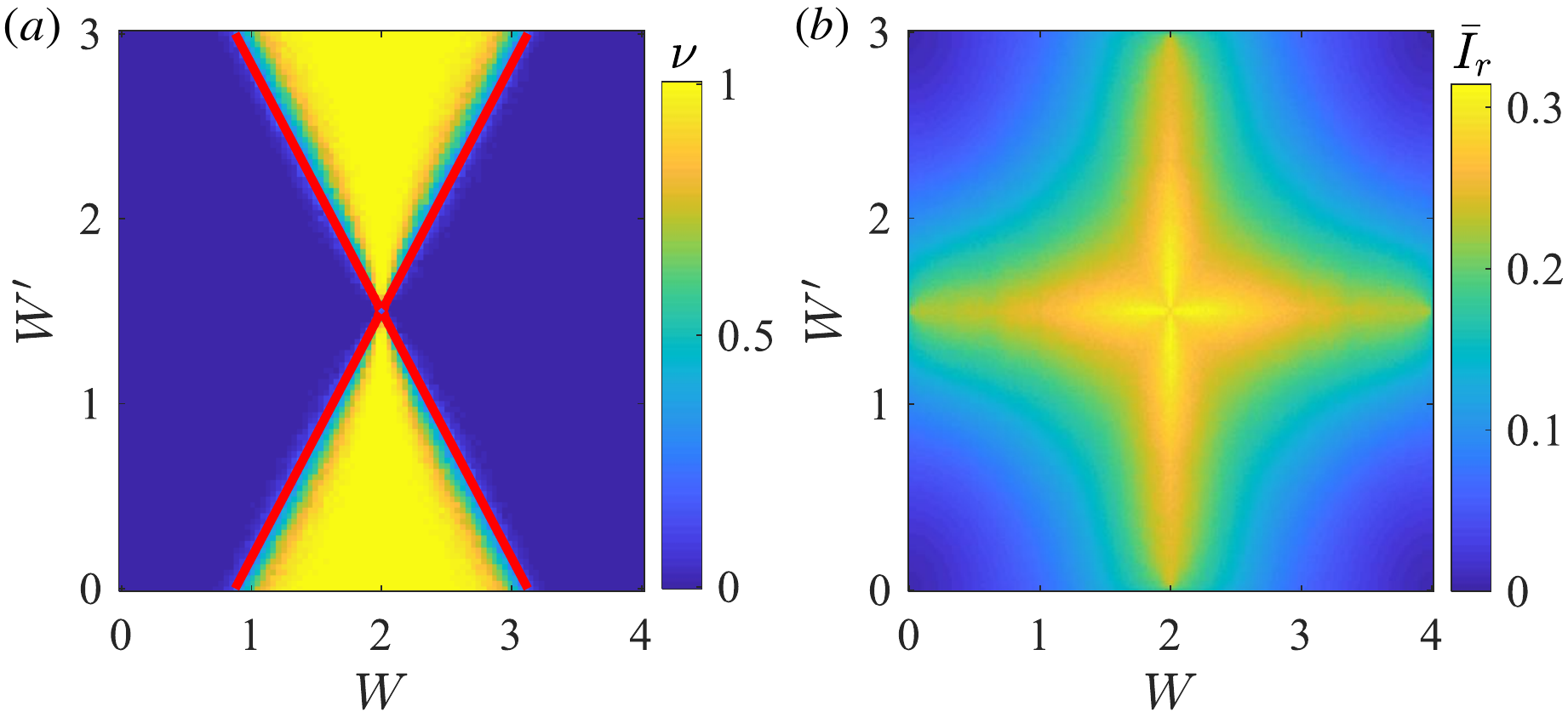}
	\caption{(Color online) \textcolor{black}{(a) The topological phase diagram in the $W$-$W'$ plane for $t=1.5$, $P=P^{\prime}=0.5$. Solid red curves indicate the topological phase boundaries revealed by the divergence of the localization length $\Lambda$. (b) The disorder-averaged real-space IPR $\bar I_r$ as a function of $W$ and $W^{\prime}$ for $t=1.5$, $P=P^{\prime}=0.5$. The lattice size is $L = 200$.}
	}\label{fig5}
\end{figure}

\section{discussion and conclusion}\label{sec5}
{\textcolor{black}{Before concluding, we discuss the 1D SSH model with overall random binary disordered hoppings and reveal the TAI phase in this general case. The intercell hoppings $t_n$ are drawn from the following bimodal probability distribution function
\begin{equation}
	\mathcal{P}^{\prime}(t_{n})=P^{\prime} \delta(t_{n}-t) + (1-P^{\prime}) \delta(t_{n}-t+W^{\prime}),
\end{equation}
where $W^{\prime}$ is the disorder strength of the intercell hopping. When $W^{\prime}=0$, it returns to the previous case discussed in Sec.~\ref{sec2}. The trivial-topological and topological-trivial phase transition points ($W \simeq 0.88$ and $W \simeq 3.12$) shown in Fig.~\ref{fig5}(a) consist with results in Fig.~\ref{fig1}(b) when $t=1.5$. When increasing the disorder strength $W^{\prime}$ of intercell hoppings, we find the TAI region shrinking and finally disappearing when $W^{\prime}=1.5$. Further increasing $W^{\prime}$ can revive and broaden the topological region. When $0<W^{\prime}<1.5$, the effective intercell hopping $\bar{t}<t=1.5$, and from the SCBA renormalized relation Eq.~(\ref{eq_crit}) one can see that the critical disorder $W_c$ from trivial to topology is larger than the corresponding point for $W'=0$ to satisfy $\bar{m}=\bar{t}$, and the TAI region induced by $W$ is shrinking. When $W'=1.5$, the probability of $t_n=0$ leads to a breaking inter-cell hopping and vanishes the TAI phase. Further increasing $W'$ effectively enlarges the hopping strength of $|t_n|$ and thus the topological phase revives. In this general case, we obtain the inverse of the localization length as
\begin{align}
\label{appeq3}
	\Lambda^{-1}=&\left|P^{\prime} \ln \right| t\left|+\left(1-P^{\prime}\right) \ln \right| t-W^{\prime}|\nonumber\\
	&-P \ln2-(1-P) \ln |2-W||
\end{align}
When $W'=0$, Eq.~(\ref{appeq3}) returns to Eq.~(\ref{appeq2}). In Fig.~\ref{fig5}(a), we plot the solid red lines where $\Lambda^{-1}=0$ on the $W$-$W'$ plane for $t=1.5$ and $P'=P=0.5$, which are consistent with the real-space winding number. We also calculate the disorder-averaged real-space IPR $\bar I_r$ in Fig.~\ref{fig5}(b), which indicates the bulk states are localized in the topological phase region.}

In summary, we have explored the topological and localization properties of the SSH model with the random binary disorder. We have shown that the random binary disorder can induce TAIs in a wide range of parameter regions of topological phase diagrams. We have analyzed trivial-topological phase transitions by calculating the renormalization of hopping strengths with the SCBA method and topological phase boundaries by calculating the localization length of the zero-energy mode. The localization and the effect of the bimodal probability have been studied. We have also numerically demonstrated that the revealed topological phases can be detected from the mean chiral displacement, which could be measured from the bulk dynamics in atomic or photonic systems. This work provides an extension of the TAIs to the case of spatially correlated disorders.

\begin{acknowledgments}
This work was supported by the National Natural Science
Foundation of China (No. 12174126, and No. 12104166), the Key-Area Research and Development Program of Guangdong Province (Grant No. 2019B030330001), the Science and Technology of Guangzhou (Grant No. 2019050001), and the Guangdong Basic and Applied Basic Research Foundation (Grants No. 2020A1515110290 and No. 2021A1515010315).

\end{acknowledgments}


%

\end{document}